\documentclass[aps,prd,reprint,nofootinbib,superscriptaddress,longbibliography,floatfix,raggedbottom]{revtex4-2}
\usepackage{amsmath,amssymb,bm,graphicx,booktabs,etoolbox,xcolor}
\allowdisplaybreaks[1]
\usepackage[T1]{fontenc}
\usepackage[utf8]{inputenc}
\usepackage[hidelinks]{hyperref}
\graphicspath{{figures/}}
\newcommand{\dd}{\mathrm{d}}
\newcommand{\im}{\operatorname{Im}}
\newcommand{\re}{\operatorname{Re}}
\newcommand{\Lb}{\Lambda_b}
\newcommand{\Lc}{\Lambda_c}
\newcommand{\cV}{\mathcal C_V}
\newcommand{\cA}{\mathcal C_A}
\newcommand{\cS}{\mathcal C_S}
\newcommand{\cP}{\mathcal C_P}
\newcommand{\cT}{\mathcal C_T}

\newcommand{\msbar}{\overline{\mathrm{MS}}}
\newcommand{\avg}[1]{\langle #1\rangle}

\begin{document}
\title{Probing new $C\!P$-violating sources in
\texorpdfstring{$\Lambda_b\to\Lambda_c\ell^-\bar\nu_\ell$}
{baryonic semileptonic decays}}
\author{Dong-Hao Li}
\email[Contact author: ]{lidh@lzu.edu.cn}
\affiliation{Frontiers Science Center for Rare Isotopes,
and School of Nuclear Science and Technology,
Lanzhou University, Lanzhou 730000, China}
\author{Meng-Chao Gao}
\thanks{Contact author: gmc8023@mails.ccnu.edu.cn}
\affiliation{Institute of Particle Physics and Key Laboratory of Quark and Lepton Physics (MOE), Central China Normal University, Wuhan 430079, China}
\author{Qin Qin}
\affiliation{School of Physics, Huazhong University of Science and Technology, Wuhan 430074, China}
\affiliation{Institute for Quantum Science and Engineering, Huazhong University of Science and Technology, Wuhan, Hubei, China}
\affiliation{Hubei Fundamental Research Center for Physics, Wuhan, Hubei, China}
\affiliation{Hubei Key Laboratory of Gravitation and Quantum Physics, Huazhong University of Science and Technology, Wuhan, Hubei, China}
\author{Meng-Han Wu}
\affiliation{School of Physics, Huazhong University of Science and Technology, Wuhan 430074, China}
\author{Fu-Sheng Yu}
\affiliation{Frontiers Science Center for Rare Isotopes,
and School of Nuclear Science and Technology,
Lanzhou University, Lanzhou 730000, China}
\affiliation{Lanzhou Center for Theoretical Physics, Key Laboratory of Theoretical Physics of Gansu Province, Key Laboratory of Quantum Theory and Applications of MoE, Gansu Provincial Research Center for Basic Disciplines of Quantum Physics, Lanzhou University, Lanzhou 730000, China}
\author{Xing-Bo Yuan}
\affiliation{Institute of Particle Physics and Key Laboratory of Quark and Lepton Physics (MOE), Central China Normal University, Wuhan 430079, China}
\begin{abstract}
While baryonic $C\!P$ violation has been observed, its potential to reveal new physics beyond the sensitivity of mesonic decays remains an open question. 
We investigate this potential within semileptonic $b\to c\ell\bar\nu_\ell$ transitions by comparing the angular distributions of $\Lambda_b\to\Lambda_c(\to\Lambda\pi)\ell^-\bar\nu_\ell$ and $\bar B\to D^*(\to D\pi)\ell^-\bar\nu_\ell$ utilizing lattice-QCD form factors.
We perform a global fit to $R(D^{(*)})$ alongside auxiliary $C\!P$-even measurements to delineate the phenomenologically viable parameter space for new-physics Wilson coefficients.
Leveraging these allowed ranges and the lattice inputs, we evaluate the complementary sensitivity of both angular distributions to novel weak phases.
We demonstrate that the baryonic decay mode probes combinations of Wilson coefficients that remain inaccessible in mesonic angular distributions, thereby providing critical supplementary information for identifying new physics.
\end{abstract}
\maketitle
\raggedbottom

\section{Introduction}
The observation of $C\!P$ violation in $\Lambda_b^0\to pK^-\pi^+\pi^-$ has extended experimental studies of this phenomenon to the baryon sector~\cite{LHCb:2025ray}. Although numerous theoretical studies have addressed this topic~\cite{Wang:2024oyi,Hsiao:2014mua,Zhu:2018jet,Geng:2021nkl,Zhang:2021fdd,Wang:2022fih,Zhang:2022iye,Rui:2022jff,Rui:2022sdc,Zhang:2022emj,Geng:2022osc,Geng:2023nia,Rui:2023fiz,Han:2024kgz,Wang:2024qff,Duan:2024zjv,Zhao:2024ren,Wang:2024rwf,Han:2025tvc,Cheng:2025kpp,Qi:2025zna,Zhang:2025mne,Sun:2025bbh,Chen:2025puj,Geng:2025yna,Shang:2026knt,Feng:2026soj,Li:2026zrs,Rui:2026ihu}, it remains unclear to what extent baryonic decays can provide insights beyond those obtained from mesonic measurements, particularly in searches for new physics (NP).
Baryon spin and parity-violating secondary decays offer access to distinct angular correlations, which depend on interference terms fundamentally different from those in meson decays~\cite{Bensalem:2002ys,Durieux:2015zwa,Wang:2022fih,Shen:2023eln,Wang:2024qff,Boer:2019zmp}.

Semileptonic $\Lambda_b\to\Lambda_c\ell^-\bar\nu_\ell$ decays offer a direct way to examine this complementarity. The underlying $b\to c$ charged-current transition is Cabibbo favored, and its leading Standard Model (SM) amplitude contains a single weak phase. 
Consequently, any nonzero $C\!P$-odd angular signal would indicate the presence of an additional weak phase. 
These decays are experimentally accessible, while their theoretical description depends on transition form factors calculated in lattice QCD, including the tensor currents needed to study general dimension-six interactions~\cite{Detmold:2015aaa,Datta:2017aue}.

The lepton-flavor universality ratios $R(D)$ and $R(D^*)$, together with polarization and kinematic measurements in $b\to c\tau\bar\nu_\tau$, constrain NP Wilson coefficients through $C\!P$-even observables~\cite{Fajfer:2012vx,Celis:2012dk,Jung:2018lfu,Murgui:2019czp,Hu:2018veh,Yan:2019hpm,Wang:2019trs,Iguro:2024hyk}. 
These constraints alone often leave the sign of a weak phase unresolved~\cite{Kumbhakar:2020jdz}. 
The angular distributions of $\bar B\to D^*(\to D\pi)\ell^-\bar\nu_\ell$ and $\Lambda_b\to\Lambda_c(\to\Lambda\pi)\ell^-\bar\nu_\ell$ have been derived in Refs.~\cite{Duraisamy:2013pia,Duraisamy:2014sna,Colangelo:2018cnj,Bhattacharya:2019olg,Becirevic:2019tpx} and~\cite{Boer:2019zmp}, respectively. Notably, the baryonic distribution encompasses scalar-current interference terms that are absent from the $\bar B\to D^*$ channel. These theoretical results enable a comprehensive comparison of the specific weak-phase combinations constrained by each decay mode. For $\bar B\to D\ell^-\bar\nu_\ell$, the spin-summed distribution contains only one polar angle and no $C\!P$-odd angular observable~\cite{Becirevic:2019tpx}. 

We employ a unified $b\to c\ell\bar\nu_\ell$ effective Hamiltonian and the angular distributions to systematically map the Wilson-coefficient interference terms in both channels. 
For $\Lambda_b\to\Lambda_c$ we use the correlated lattice-QCD vector, axial-vector, and tensor form factors~\cite{Detmold:2015aaa,Datta:2017aue}. For $\bar B\to D^*$ we adopt the HPQCD lattice form factors parameterized via a Boyd-Grinstein-Lebed (BGL) expansion subject to weak unitarity bounds~\cite{Harrison:2023dzh,Boyd:1997kz,Bordone:2026gqs}. This consistent phenomenological framework facilitates a direct comparison of angular observables inclusive of their respective hadronic uncertainties.

We fit $R(D^{(*)})$ and other measured $C\!P$-even $b\to c\tau\bar\nu_\tau$ observables to constrain one complex Wilson coefficient at a time in both the chiral and parity bases. We subsequently utilize these allowed regions in conjunction with the lattice form factors to compute the $C\!P$-odd angular moments in both the mesonic and baryonic decay channels. Our comparison reveals that the baryonic moments can cleanly resolve a scalar weak-phase combination to which the $\bar B\to D^*$ angular distribution is completely insensitive, thereby supplying vital information beyond existing mesonic and $C\!P$-even measurements. The angular-distribution framework can be also applied to electron and muon final states, with the corresponding lepton masses and flavor-specific Wilson coefficients.

\section{Effective Hamiltonian and operator basis}
\label{sec:eft}
The effective Hamiltonian establishes the Wilson-coefficient conventions utilized in both our numerical fits and the derivation of the angular distributions. We first define the dimension-six operators in the chiral basis and subsequently group their coefficients according to the parity of the hadronic currents. The resulting combinations isolate the weak-phase interference effects compared in subsequent sections.

For each lepton flavor and a left-handed neutrino, the generalized effective Hamiltonian is written as
\begin{align}
 {\cal H}_{\rm eff}=&\frac{4G_FV_{cb}}{\sqrt2}
 \big[(1+C_{V_L})O_{V_L}+C_{V_R}O_{V_R}
 \nonumber\\
&+C_{S_L}O_{S_L}+C_{S_R}O_{S_R}+2C_TO_T\big] +\text{h.c.},\label{eq:ham}
\end{align}
with the operators
\begin{align}
 O_{V_{L,R}}&=(\bar c\gamma^\mu P_{L,R}b)(\bar\ell\gamma_\mu P_L\nu_\ell),\nonumber\\
 O_{S_{L,R}}&=(\bar cP_{L,R}b)(\bar\ell P_L\nu_\ell),\nonumber\\
 O_T&=(\bar c\sigma^{\mu\nu}P_Lb)
(\bar\ell\sigma_{\mu\nu}P_L\nu_\ell),\nonumber
\end{align}
where $P_{L,R}=(1\mp\gamma_5)/2$ and $\sigma^{\mu\nu}=i[\gamma^\mu,\gamma^\nu]/2$.
We use a single tensor coefficient $C_T$, normalized to the tensor quark current in the parity basis with the leptonic current kept chiral. The coefficient multiplying the chiral operator $O_T$ is therefore $2C_T$. Explicitly,
\begin{align*}
 2C_TO_T={}&C_T\big[(\bar c\sigma^{\mu\nu}b)
 -(\bar c\sigma^{\mu\nu}\gamma_5b)\big]\times(\bar\ell\sigma_{\mu\nu}P_L\nu_\ell).
\end{align*}
Thus the tensor and pseudotensor quark currents carry coefficients $C_T$ and $-C_T$, respectively, with the common prefactor $4G_FV_{cb}/\sqrt2$ understood.
The coefficients $C_i\equiv C_i^\ell$ describe flavor-dependent NP contributions and vanish in the SM.  We suppress the lepton-flavor label and define all amplitudes at a common low scale $\mu_b\sim m_b$.  
The angular amplitudes use the unit SM coefficient in Eq.~\eqref{eq:ham}, and the rate-fit normalization is specified in Sec.~\ref{sec:numerics}. In our benchmark scenarios, NP contributions are assigned exclusively to the tau channel.

For the angular amplitudes, we define the total coefficients as
\begin{align}
 \cV&=1+C_{V_L}+C_{V_R},&
 \cA&=1+C_{V_L}-C_{V_R},\nonumber\\
 \cS&=C_{S_L}+C_{S_R},&
 \cP&=C_{S_R}-C_{S_L},\nonumber\\
 \cT&=2C_T.&& \label{eq:parity}
\end{align}
The quark currents can then be equivalently parameterized as $\bar c\gamma^\mu(\cV-\cA\gamma_5)b$ and $\bar c(\cS+\cP\gamma_5)b$. The tensor amplitudes use $\cT=2C_T$, the coefficient of $O_T$ in Eq.~\eqref{eq:ham}. These combinations specify the short-distance factors in all angular expressions below.

\section{Lattice form factors and weak unitarity bound}
\label{sec:ff}
We use state-of-the-art lattice-QCD transition form factors to evaluate the angular distributions of the two decay channels.
The $\Lb\to\Lc$ input includes vector, axial-vector, and tensor currents with their correlations. The $\bar B\to D^*$ input includes the corresponding currents in a weak-unitarity-constrained parameterization. These inputs determine the hadronic amplitudes and their associated uncertainties.
Earlier studies of $\Lb\to\Lc\tau\bar\nu_\tau$ examined its angular and polarization observables in the SM and with NP interactions, including leptoquark scenarios~\cite{Shivashankara:2015cta,Gutsche:2015mxa,Li:2016pdv}. Heavy-quark symmetry gives complementary constraints on the baryonic form factors~\cite{Bernlochner:2018bfn}.

\subsection{\texorpdfstring{$\Lb\to\Lc$}{Lambda b to Lambda c}}
We implement the full set of ten helicity form factors:
\begin{equation}
\{f_+,f_0,f_\perp,g_+,g_0,g_\perp,
h_+,h_\perp,\widetilde h_+,\widetilde h_\perp\}
\end{equation}
following Refs.~\cite{Detmold:2015aaa,Datta:2017aue}.
For clarity, the defining matrix elements are given here. Let $p$ and $p'$ be the $\Lb$ and $\Lc$ momenta, $q=p-p'$, and $s_\pm=(m_{\Lb}\pm m_{\Lc})^2-q^2$.

We also define the auxiliary vector
\begin{equation}
P_q^\mu=p^\mu+p'^\mu-\frac{m_{\Lb}^2-m_{\Lc}^2}{q^2}q^\mu .
\end{equation}
With spinors $u_b=u_{\Lb}(p)$ and $\bar u_c=\bar u_{\Lc}(p')$, the vector and axial-vector matrix elements are parameterized as
\begin{align}
 &\langle\Lc|\bar c\gamma^\mu b|\Lb\rangle\nonumber\\
 &=\bar u_c\bigg[f_0(m_{\Lb}-m_{\Lc})\frac{q^\mu}{q^2}
       +f_+\frac{m_{\Lb}+m_{\Lc}}{s_+}P_q^\mu \nonumber\\
 &\quad
    +f_\perp\left(\gamma^\mu-\frac{2m_{\Lc}}{s_+}p^\mu
    -\frac{2m_{\Lb}}{s_+}p'^\mu\right)\bigg]u_b ,
 \label{eq:baryon-vector}\\
 &\langle\Lc|\bar c\gamma^\mu\gamma_5 b|\Lb\rangle\nonumber\\
 &=-\bar u_c\gamma_5\bigg[g_0(m_{\Lb}+m_{\Lc})\frac{q^\mu}{q^2}
       +g_+\frac{m_{\Lb}-m_{\Lc}}{s_-}P_q^\mu \nonumber\\
 &\quad
+g_\perp\left(\gamma^\mu+\frac{2m_{\Lc}}{s_-}p^\mu
-\frac{2m_{\Lb}}{s_-}p'^\mu\right)\bigg]u_b .
 \label{eq:baryon-axial}
\end{align}
To specify the tensor convention compactly, define $D^{\mu\nu}=p^\mu p'^\nu-p^\nu p'^\mu$ and $Q^{\mu\nu}(k)=k^\mu\gamma^\nu-k^\nu\gamma^\mu$.
The uncontracted tensor matrix element of Ref.~\cite{Datta:2017aue} is
\begin{widetext}
\begin{align}
 &\langle\Lc|\bar c\,i\sigma^{\mu\nu}b|\Lb\rangle
 =\bar u_c\bigg\{
 \frac{2h_+}{s_+}D^{\mu\nu}
 +h_\perp\left[\frac{m_{\Lb}+m_{\Lc}}{q^2}Q^{\mu\nu}(q)
-2\left(\frac1{q^2}+\frac1{s_+}\right)D^{\mu\nu}\right]
 \nonumber\\
 &+\widetilde h_+\left[i\sigma^{\mu\nu}
       -\frac2{s_-}\left(m_{\Lb}Q^{\mu\nu}(p')-m_{\Lc}Q^{\mu\nu}(p)+D^{\mu\nu}\right)\right]
 \nonumber\\
 &+\widetilde h_\perp\frac{m_{\Lb}-m_{\Lc}}{q^2s_-}
 \left[-(m_{\Lb}^2-m_{\Lc}^2-q^2)Q^{\mu\nu}(p)
       +(m_{\Lb}^2-m_{\Lc}^2+q^2)Q^{\mu\nu}(p')
       +2(m_{\Lb}-m_{\Lc})D^{\mu\nu}\right]\bigg\}u_b .
 \label{eq:baryon-tensor}
\end{align}
\end{widetext}
We take $g_{\mu\nu}=\operatorname{diag}(1,-1,-1,-1)$, $\epsilon_{0123}=+1$, and $\sigma^{\mu\nu}\gamma_5=-i\epsilon^{\mu\nu\alpha\beta}\sigma_{\alpha\beta}/2$. The identity fixes the matrix element of pseudotensor current $\bar c\sigma^{\mu\nu}\gamma_5 b$ without additional form factors.

The primary six vector and axial-vector functions originate from the $n_f=2+1$ RBC/UKQCD-ensemble computations employing relativistic heavy quarks, as detailed in Ref.~\cite{Detmold:2015aaa}. Reference~\cite{Datta:2017aue} extended this basis by incorporating the four tensor functions, supplying the fully correlated ten-form-factor global fit implemented herein.
In the physical limit, the lattice fit uses the first-order $z$ expansion~\cite{Detmold:2015aaa,Datta:2017aue}
\begin{align}
 F(q^2)&=\frac{a_0^F+a_1^F z_F(q^2)}{1-q^2/(m_{\rm pole}^F)^2},\label{eq:lbz}\\
 z_F(q^2)&=
 \frac{\sqrt{t_+^F-q^2}-\sqrt{t_+^F-t_-}}{\sqrt{t_+^F-q^2}+\sqrt{t_+^F-t_-}},\nonumber\\
 t_-&=(m_{\Lb}-m_{\Lc})^2,\qquad
 t_+^F=(m_{\rm pole}^F)^2.\nonumber
\end{align}
Following Refs.~\cite{Detmold:2015aaa,Datta:2017aue}, we set the mapping parameter $t_+^F$ equal to the squared pole mass for each baryonic form factor. The physical $BD$ branch cut begins at $(m_B+m_D)^2$. The mesonic BGL expansion below uses the $BD^*$ pair-production threshold. The pole masses in GeV are
\begin{equation*}
\begin{array}{c|ccccc}
F&f_+,f_\perp,h_+,h_\perp&f_0&g_+,g_\perp&
 g_0&\widetilde h_+,\widetilde h_\perp\\ \hline
 m_{\rm pole}^F&6.332&6.725&6.768&6.276&6.768
\end{array}
\end{equation*}
We use the baryon masses $m_{\Lb}=5.61960$ GeV and $m_{\Lc}=2.28646$ GeV.
The constraints $g_\perp(t_-)=g_+(t_-)$ and $\widetilde h_\perp(t_-)=\widetilde h_+(t_-)$ rigorously identify the corresponding constant coefficients, thereby reducing the system to 18 independent fit parameters.

The statistical fit incorporates the central values and the complete covariance matrix documented in Ref.~\cite{Datta:2017aue}. Scalar and pseudoscalar matrix elements are uniquely determined via the equations of motion, circumventing the need for additional independent form factors:
\begin{align}
 q_\mu\langle\bar c\gamma^\mu b\rangle
   &=(m_b-m_c)\langle\bar c b\rangle,\\
 q_\mu\langle\bar c\gamma^\mu\gamma_5 b\rangle
   &=-(m_b+m_c)\langle\bar c\gamma_5 b\rangle.\nonumber
\end{align}
It is mandatory that quark masses and corresponding scalar coefficients be defined within an identical renormalization scheme and evaluated at the same energy scale. The tensor form factors extracted from Ref.~\cite{Datta:2017aue} are specifically provided in the $\msbar$ scheme at $\mu=m_b$.

\begin{figure*}[t]
\centering
\includegraphics[width=.82\textwidth]{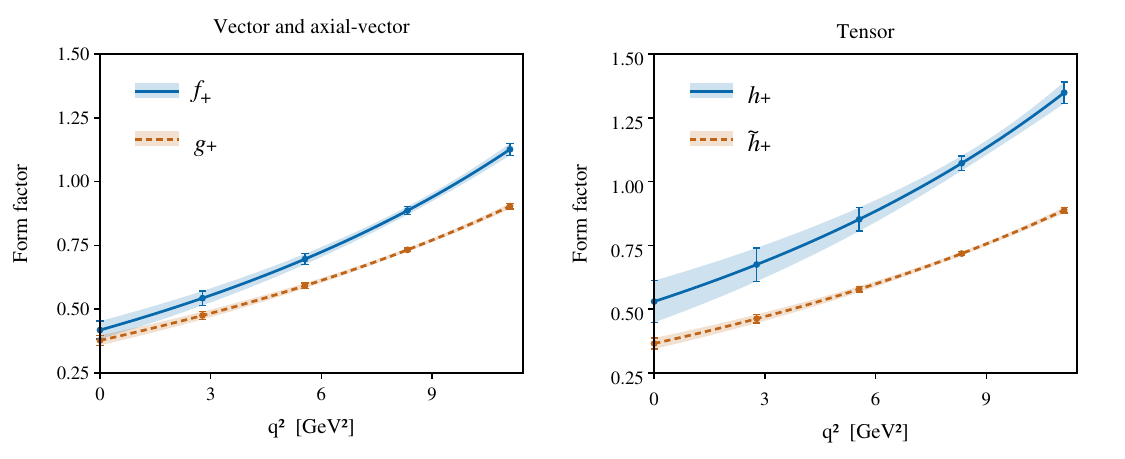}
\caption{Representative $\Lambda_b\to\Lambda_c$ form factors from the nominal lattice fit~\cite{Datta:2017aue}: $f_+$ and $g_+$ (left), and $h_+$ and $\widetilde h_+$ (right). Curves and bands show Eq.~\eqref{eq:lbz} and its one-standard-deviation uncertainty. Points with error bars are correlated evaluations of the same fit at five values of $q^2$.}
\label{fig:baryon-ff}
\end{figure*}

The aforementioned lattice analyses also incorporate higher-order fits to rigorously quantify systematic uncertainties stemming from the $z$ expansion truncation, continuum and chiral extrapolations, and requisite current matching procedures.
For the numerical benchmarks we use the covariance of the preferred first-order $z$-expansion in Refs.~\cite{Detmold:2015aaa,Datta:2017aue}.
Fig.~\ref{fig:baryon-ff} shows $f_+$ and $g_+$ together, and the representative tensor form factors $h_+$ and $\widetilde h_+$ in the second panel, with uncertainties propagated from this covariance.

\subsection{\texorpdfstring{$\bar B\to D^*$}{B to D star}}
Calculations by Fermilab Lattice/MILC and JLQCD also determine $\bar B\to D^*$ form factors beyond zero recoil~\cite{FermilabLattice:2021cdg,Aoki:2023qpa}. The HPQCD calculation~\cite{Harrison:2023dzh} determines the vector, axial-vector, and tensor form factors over the full semileptonic range using MILC $n_f=2+1+1$ HISQ ensembles, the heavy-HISQ method, and nonperturbatively normalized currents.  
Three bases enter at different stages of our analysis. The angular amplitudes are evaluated in the phenomenological basis
\begin{equation}
 \{V,A_0,A_1,A_{12},T_1,T_2,T_{23}\}.
 \label{eq:amplitude-basis}
\end{equation}
HPQCD supplies its physical-continuum posterior in the HQET basis
$\{h_{A_1},h_{A_2},h_{A_3},h_V,h_{T_1},h_{T_2},h_{T_3}\}$.
We first convert those correlated results to Eq.~\eqref{eq:amplitude-basis}, then perform the unitarity-constrained fit in the BGL basis $\{f,{\cal F}_1,{\cal F}_2,g,T_1,T_2,T_{23}\}$, and finally convert the BGL posterior back to Eq.~\eqref{eq:amplitude-basis} for all angular predictions. This ordering keeps the lattice input, the dispersive bounds, and the amplitude conventions conceptually separate.
Let $p$ and $k$ denote the $\bar B$ and $D^*$ momenta, $q=p-k$, and $\varepsilon$ the $D^*$ polarization. The vector and axial-vector matrix elements are
\begin{align}
 &\langle D^*|\bar c\gamma_\mu b|\bar B\rangle=-i\epsilon_{\mu\nu\alpha\beta}
    \varepsilon^{*\nu}p^\alpha k^\beta\frac{2V}{m_B+m_{D^*}},
 \nonumber\\
 &\langle D^*|\bar c\gamma_\mu\gamma_5 b|\bar B\rangle\nonumber\\
 &=(m_B+m_{D^*})A_1\varepsilon_\mu^*
-\frac{A_2}{m_B+m_{D^*}}(\varepsilon^*\cdot q)(p+k)_\mu
   \nonumber\\
 &-\frac{2m_{D^*}}{q^2}(\varepsilon^*\cdot q)q_\mu(A_3-A_0),
 \label{eq:meson-axial}\\
 A_3&=\frac{m_B+m_{D^*}}{2m_{D^*}}A_1-\frac{m_B-m_{D^*}}{2m_{D^*}}A_2 .\nonumber
\end{align}
The corresponding scalar and pseudoscalar matrix elements are subsequently extracted as
\begin{align}
 \langle D^*|\bar c b|\bar B\rangle&=0,\nonumber\\
 \langle D^*|\bar c\gamma_5 b|\bar B\rangle
 &=-\frac{2m_{D^*}}{m_b+m_c}(\varepsilon^*\cdot q)A_0 .
 \label{eq:meson-scalar}
\end{align}
The vanishing scalar matrix element governs the scalar sensitivity discussed in Sec.~\ref{sec:complementarity}.
The definition of tensor form factor reads
\begin{align}
 \langle D^*|\bar c\sigma_{\mu\nu}b|\bar B\rangle
&=\epsilon_{\mu\nu\alpha\beta}\big[
 \varepsilon^{*\alpha}(p+k)^\beta g_+\nonumber\\
&\quad+\varepsilon^{*\alpha}q^\beta g_-
 +(\varepsilon^*\cdot q)p^\alpha k^\beta g_0\big],
 \label{eq:meson-tensor}
\end{align}
with
\begin{align}
 g_+&=-T_1,\qquad
 g_-=\frac{m_B^2-m_{D^*}^2}{q^2}(T_1-T_2),\nonumber\\
 g_0&=\frac2{q^2}\left[T_1-T_2-\frac{q^2}{m_B^2-m_{D^*}^2}T_3\right].
\end{align}
The pseudotensor follows from the same Dirac identity as in the baryonic channel.
The kinematic relations $A_0(0)=A_3(0)$ and $T_1(0)=T_2(0)$ remove the apparent $q^2=0$ poles in these decompositions.

Writing $r=m_{D^*}/m_B$ and $w=(m_B^2+m_{D^*}^2-q^2)/(2m_B m_{D^*})$,
we convert to the phenomenological basis through
\begin{align}
 V&=\frac{1+r}{2\sqrt r}h_V,\qquad
 A_1=\frac{\sqrt r(1+w)}{1+r}h_{A_1},\nonumber\\
 A_0&=\frac{(1+w)h_{A_1}+(rw-1)h_{A_2}
+(r-w)h_{A_3}}{2\sqrt r},\nonumber\\
 A_2&=\frac{1+r}{2\sqrt r}(rh_{A_2}+h_{A_3}),\nonumber\\
 T_1&=\frac{(1+r)h_{T_1}-(1-r)h_{T_2}}{2\sqrt r},\nonumber\\
 T_2&=\sqrt r\left[\frac{w+1}{1+r}h_{T_1}
-\frac{w-1}{1-r}h_{T_2}\right],\nonumber\\
 T_3&=\frac{(1-r)h_{T_1}-(1+r)h_{T_2}
+(1-r^2)h_{T_3}}{2\sqrt r}.
 \label{eq:hpqcd-dictionary}
\end{align}
We propagate the complete HPQCD covariance through this change of basis.
The HPQCD tensor input is renormalized in the $\msbar$ scheme at $\mu=4.8$ GeV~\cite{Harrison:2023dzh}. The physical-continuum curves and their covariance are taken from the ancillary material of Ref.~\cite{Harrison:2023dzh}.  We evaluate them at the synthetic kinematic points used in the BGL analysis and retain the resulting correlations.
With $\lambda=(m_B^2+m_{D^*}^2-q^2)^2-4m_B^2m_{D^*}^2$, the two helicity combinations used in Eq.~\eqref{eq:amplitude-basis} are
\begin{align}
 A_{12}&=
 \frac{(m_B+m_{D^*})^2(m_B^2-m_{D^*}^2-q^2)A_1-\lambda A_2}
 {16m_Bm_{D^*}^2(m_B+m_{D^*})},\nonumber\\
 T_{23}&=
 \frac{(m_B^2-m_{D^*}^2)(m_B^2+3m_{D^*}^2-q^2)T_2-\lambda T_3}
 {8m_Bm_{D^*}^2(m_B-m_{D^*})}.
\end{align}
The corresponding BGL basis is
\begin{align}
 f&=(m_B+m_{D^*})A_1,& {\cal F}_1&=8m_Bm_{D^*} A_{12},\nonumber\\
 {\cal F}_2&=2A_0,& g&=\frac{2V}{m_B+m_{D^*}},
\end{align}
together with $T_1,T_2,T_{23}$. The correlated HPQCD form factors are evaluated with $m_B=5.27964$ GeV and $m_{D^*}=2.010$ GeV.

For each of these seven functions,
\begin{align}
 F_i(q^2)&=\frac{1}{P_i(z)\phi_i(z)}
\sum_{n=0}^{2}a_n^{(i)}z^n,\label{eq:bgl}\\
 z(q^2)&=\frac{\sqrt{t_+-q^2}-\sqrt{t_+-t_0}}{\sqrt{t_+-q^2}+\sqrt{t_+-t_0}},\nonumber\\
 t_\pm&=(m_B\pm m_{D^*})^2,\quad t_0=t_-.\nonumber
\end{align}
Here $P_i$ removes subthreshold poles and $\phi_i$ is the outer function normalized with the appropriate current susceptibility~\cite{Boyd:1997kz,Bordone:2026gqs}.
We truncate the expansion at $n=2$, with three coefficients for each form factor.
The 21 coefficients obey the four linear relations implemented in the fit:
\begin{align}
 {\cal F}_1(t_-)&=(m_B-m_{D^*})f(t_-),\,\,
 {\cal F}_2(0)=\frac{2{\cal F}_1(0)}{m_B^2-m_{D^*}^2},\nonumber\\
 T_1(0)&=T_2(0),\,\,
 T_{23}(t_-)=\frac{(m_B+m_{D^*})^2}{4m_Bm_{D^*}}T_2(t_-).\label{eq:endpoints}
\end{align}
These are enforced by parametrizing the coefficient vector in the null space of the constraint matrix, leaving 17 fit coordinates.

\begin{figure*}[t]
\centering
\includegraphics[width=.82\textwidth]{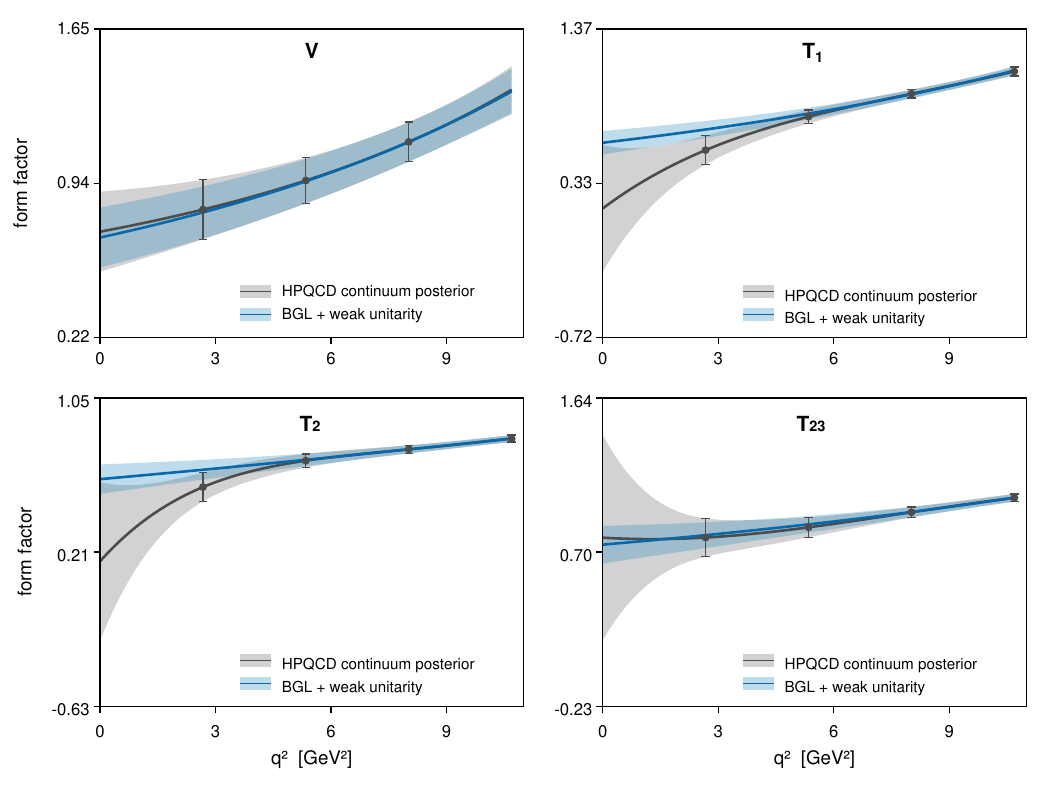}
\caption{HPQCD physical-continuum means and one-standard-deviation bands (gray), compared with the $n=2$ BGL posterior after all four kinematic constraints and weak-unitarity constraints (blue). Gray points with error bars mark the retained synthetic inputs: three interior points for $V$ and four points for each tensor form factor.    
The vector panel is representative of the small changes found for $V,A_0,A_1,A_{12}$, whereas the tensor panels display the pronounced reduction of the low-$q^2$ uncertainty.}
\label{fig:bgl-unitarity}
\end{figure*}

We construct correlated synthetic form-factor values at five recoil points across the semileptonic range.  Following the stability strategy of Ref.~\cite{Bordone:2026gqs}, the fit uses the three interior points for $f,{\cal F}_1,{\cal F}_2,g$ (discarding the two endpoints) and four points for $T_1,T_2,T_{23}$ (discarding the $q^2=0$ point). Fig.~\ref{fig:bgl-unitarity} shows the retained BGL-fit input points and the form factors obtained after imposing weak unitarity. 

With the outer-function normalization of Ref.~\cite{Bordone:2026gqs}, the weak-unitarity bounds are
\begin{align}
 \sum_{n=0}^{2}\big[(a_n^f)^2+
(a_n^{{\cal F}_1})^2\big]&\leq1,&
 \sum_{n=0}^{2}(a_n^{{\cal F}_2})^2&\leq1,\nonumber\\
 \sum_{n=0}^{2}(a_n^g)^2&\leq1,&
 \sum_{n=0}^{2}(a_n^{T_1})^2&\leq1,\nonumber\\
 \sum_{n=0}^{2}\big[(a_n^{T_2})^2+
(a_n^{T_{23}})^2\big]&\leq1.
 \label{eq:weak}
\end{align}
After the four endpoint relations have been imposed, we write $\bm\theta$ denotes the 17 independent BGL coefficients. Let $\bm y$ collect the retained correlated HPQCD values and let $\bm F_{\rm BGL}(\bm\theta)$ denote the corresponding BGL predictions. The posterior can be written as
\begin{align}
 p(\bm\theta\mid\bm y)&\propto
 e^{-\chi^2_{\rm lat}(\bm\theta)/2}
 \prod_{s=1}^{5}\Theta\!\left[1-U_s(\bm\theta)\right],\nonumber\\
 \chi^2_{\rm lat}(\bm\theta)&=
 [\bm y-\bm F_{\rm BGL}(\bm\theta)]^T
 \Sigma_y^+
 [\bm y-\bm F_{\rm BGL}(\bm\theta)].
 \label{eq:post}
\end{align}
Here $\Theta$ is the Heaviside function, $\Sigma_y^+$ is the inverse on the retained covariance subspace, and the five $U_s$ are the sums on the left-hand sides of Eq.~\eqref{eq:weak}. 
We first obtain the least-squares solution $\hat{\bm \theta}$ and its covariance $C_\theta$ from $\chi_{\text{lat}}^2$, draw candidate points from $(\hat{\bm\theta},C_\theta)$, and retain $10^4$ accepted points satisfying all five unitarity bounds. 
This rejection procedure samples the posterior in Eq.~\eqref{eq:post}, corresponding to a constant prior on the independent BGL coefficients together with the five hard unitarity bounds.
Imposing the strong unitarity bounds requires additional HQET parameters and is beyond the scope of this work.
After the weak-unitarity-constrained BGL fit, the vector form factor $V$, shown as a representative example in Fig.~\ref{fig:bgl-unitarity}, remains close to the HPQCD continuum result. The same holds for $A_0,A_1,A_{12}$. 
In contrast, the HPQCD continuum result constrains $T_1,T_2,T_{23}$ only weakly at low $q^2$, and the unitarity-constrained BGL posterior substantially reduces their uncertainties in that region.

\section{Angular distributions and weak-phase interference}
\label{sec:angular}
We use the angular distributions derived for the mesonic and baryonic decays in Refs.~\cite{Bhattacharya:2019olg,Becirevic:2019tpx,Boer:2019zmp}. For the mesonic channel, we have further verified that our implementation reproduces the known relations among angular coefficients in the corresponding limits discussed in Ref.~\cite{Alguero:2020ukk}.
In the common operator basis of Sec.~\ref{sec:eft}, their $C\!P$-sensitive terms depend on different imaginary Wilson-coefficient interferences. We specify the angular conventions and the terms needed for the comparison. The form-factor expressions for the amplitudes are collected in the Appendix.

\begin{figure*}[t]
\centering
\includegraphics[width=.63\textwidth]{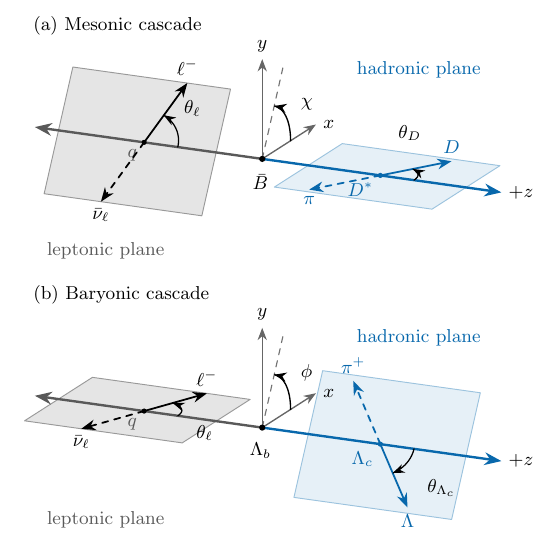}
\caption{Decay planes and angles for (a) $\bar B\to D^*(\to D\pi)\ell^-\bar\nu_\ell$ and (b) $\Lb\to\Lc(\to\Lambda\pi^+)\ell^-\bar\nu_\ell$. The gray leptonic and blue hadronic plane patches are separated for clarity. Their longitudinal axes meet at the parent vertex. The $+z$ axis follows the daughter hadron. Polar angles are defined in the corresponding daughter-hadron or dilepton rest frame. The $x$ and $y$ axes specify the azimuths directly: $\chi$ runs from the $D$ transverse direction to that of $\ell^-$, and $\phi$ from the $\ell^-$ transverse direction to that of $\pi^+$, about $+z$.}
\label{fig:decay-angles}
\end{figure*}

We neglect absorptive phases in the form factors and use the narrow-width approximation for the daughter-hadron decays.
The angular conventions for both cascades are illustrated in Fig.~\ref{fig:decay-angles}.

\subsection{\texorpdfstring{$\bar B\to D^*$}{B to D star}}\label{sec:meson}
We follow the angular and amplitude conventions of Refs.~\cite{Bhattacharya:2019olg,Becirevic:2019tpx}. The $C\!P$-sensitive mesonic coefficients receive vector, axial-vector, pseudoscalar, and tensor contributions but are independent of the scalar current.

Following Ref.~\cite{Becirevic:2019tpx}, the invariant $q^2=(p_\ell+p_\nu)^2$ is the dilepton mass squared. In Fig.~\ref{fig:decay-angles}(a), the $+z$ axis follows the $D^*$ in the $\bar B$ rest frame. The angles $\theta_D$ and $\theta_\ell$ measure the $D$ and charged-lepton directions relative to this axis in the $D^*$ and dilepton rest frames. Choosing $\bm p_D$ in the $xz$ plane with positive $x$ component, $\chi$ is the charged-lepton azimuth about $+z$. The helicity-amplitude conventions are specified in Appendix~\ref{app:meson-amplitudes}.
The full differential distribution is
\begin{widetext}
\begin{align}
  \frac{\dd^4\Gamma}{\dd q^2\dd\cos\theta_\ell \dd\cos\theta_D\dd\chi}
 =\frac{9}{32\pi}\big\{
 &
I_{1c}\cos^2\theta_D+I_{1s}\sin^2\theta_D\nonumber\\
 &+(I_{2c}\cos^2\theta_D+I_{2s}\sin^2\theta_D)
 \cos2\theta_\ell\nonumber\\
 &+(I_{6c}\cos^2\theta_D+I_{6s}\sin^2\theta_D)
  \cos\theta_\ell\nonumber\\
 &+(I_3\cos2\chi+I_9\sin2\chi)
\sin^2\theta_\ell\sin^2\theta_D\nonumber\\
 &+(I_4\cos\chi+I_8\sin\chi)
\sin2\theta_\ell\sin2\theta_D\nonumber\\
 &+(I_5\cos\chi+I_7\sin\chi)
\sin\theta_\ell\sin2\theta_D\big\}.
 \label{eq:mesondist}
\end{align}
\end{widetext}
In particular,
\begin{equation}
 \frac{\dd\Gamma_M}{\dd q^2}=\frac14(3I_{1c}+6I_{1s}-I_{2c}-2I_{2s}).
\end{equation}
The moments discussed here are $\widehat I_i=I_i/(\dd\Gamma_M/\dd q^2)$ and the integrated ratios $\avg{I_i}=\int\dd q^2\,I_i(q^2)/\Gamma_M$, with $\Gamma_M=\int\dd q^2\,(\dd\Gamma_M/\dd q^2)$ over the chosen lepton's physical range. The amplitudes $H_{\pm,0,t}$ describe the transverse, longitudinal, and timelike vector/axial helicities, $H_P$ is the pseudoscalar amplitude, and $H_{T,\pm,0}$ are the tensor amplitudes.  Their form-factor expressions, the lepton-mass-dressed combinations $H_\lambda^\pm$ and $\widetilde H_t$, and the rate are given in Eqs.~\eqref{eq:mesonamps}--\eqref{eq:mesonrate}. After introducing $r_\ell=m_\ell/\sqrt{q^2}$, the $C\!P$-sensitive coefficients are
\begin{align}
 I_7={}&2{\cal N}_M\im\!\left[
 (H_+^-+H_-^-)H_0^{-*}
 -r_\ell^2(H_+^+-H_-^+)\widetilde H_t^*\right],
 \nonumber\\
 I_8={}&{\cal N}_M\im\!\left[
 (H_+^--H_-^-)H_0^{-*}
 -r_\ell^2(H_+^+-H_-^+)H_0^{+*}\right],\nonumber\\
 I_9={}&-2{\cal N}_M\im\!\left[
 H_+^-H_-^{-*}-r_\ell^2H_+^+H_-^{+*}\right].
 \label{eq:i9}
\end{align}
The positive factor ${\cal N}_M(q^2)$ is given explicitly in Eq.~\eqref{eq:mesonnorm}. Its $q^2$-independent normalization cancels in the integrated ratios $\avg{I_i}$.

Angular momentum and parity allow the pseudoscalar current, with $J^P=0^-$, to connect the $0^-$ $\bar B$ and $1^-$ $D^*$ states. The scalar $0^+$ current has a vanishing matrix element. The $I_i$ therefore depend on $\cP$ and are independent of $\cS$.
For real tensor form factors and one tensor coefficient, tensor--tensor products are real. The resulting interference terms are
\begin{align}
 I_7={}&F_{7,VA}\im(\cV\cA^*)
       +F_{7,VP}\im(\cV\cP^*)\nonumber\\
      &+F_{7,AT}\im(\cA\cT^*)
 +F_{7,PT}\im(\cP\cT^*),\nonumber\\
 I_8={}&F_{8,VA}\im(\cV\cA^*),\nonumber\\
 I_9={}&F_{9,VA}\im(\cV\cA^*).\label{eq:imap}
\end{align}
Here the real functions $F_{i,XY}(q^2)$ contain only masses, kinematics, and form factors. Thus each displayed term is directly proportional to the indicated imaginary Wilson-coefficient bilinear.
At fixed $q^2$, the three mesonic coefficients span at most two independent short-distance directions: $I_8$ and $I_9$ both measure only $\im(\cV\cA^*)$, with different kinematic weights, while $I_7$ supplies one additional weighted linear combination of $\im(\cV\cA^*)$, $\im(\cV\cP^*)$, $\im(\cA\cT^*)$, and $\im(\cP\cT^*)$.

\subsection{\texorpdfstring{$\Lb\to\Lc(\to\Lambda\pi)$}{Lambda b to Lambda c}}
\label{sec:baryon}
The comprehensive $\Lb\to\Lc(\to\Lambda\pi)\ell^-\bar\nu_\ell$ phase-space distribution was established in Ref.~\cite{Boer:2019zmp}.
The scalar matrix element of the primary $\Lb\to\Lc$ transition is nonzero and can therefore interfere with vector amplitudes in the $\Lc$ spin-density matrix.  The parity-violating secondary decay analyzes the produced $\Lc$ polarization and makes the associated interference observable through the daughter-hadron angles.  This supplies the baryonic counterpart to the mesonic coefficients above.

We assume an unpolarized $\Lb$ and use the angles of Ref.~\cite{Boer:2019zmp}: in the $\Lb$ rest frame the $\Lc$ moves along the positive $z$ axis, $\theta_{\Lc}$ measures the $\Lambda$ direction in the $\Lc$ rest frame, and $\theta_\ell$ is the lepton angle relative to this axis in the dilepton frame.
Choosing the charged lepton in the $xz$ plane with positive $x$ component, $\phi$ is the azimuth of the secondary $\pi^+$ about $+z$, as in Fig.~\ref{fig:decay-angles}(b). The transverse $\Lambda$ momentum points opposite to that of the $\pi^+$.
We use normalized coefficients $K_i$ throughout the angular
distribution:
\begin{widetext}
\begin{align}
 \frac{1}{\dd\Gamma_B/\dd q^2}
 \frac{\dd^4\Gamma}{\dd q^2\dd\cos\theta_\ell
    \dd\cos\theta_{\Lc}\dd\phi}
 =\frac{3}{8\pi}\big\{&
 K_{1ss}\sin^2\theta_\ell+
 K_{1cc}\cos^2\theta_\ell+
 K_{1c}\cos\theta_\ell\nonumber\\
 &+(K_{2ss}\sin^2\theta_\ell+
 K_{2cc}\cos^2\theta_\ell+
K_{2c}\cos\theta_\ell)\cos\theta_{\Lc}\nonumber\\
 &+(K_{3sc}\sin\theta_\ell\cos\theta_\ell+
K_{3s}\sin\theta_\ell)\sin\theta_{\Lc}\sin\phi\nonumber\\
 &+(K_{4sc}\sin\theta_\ell\cos\theta_\ell+
K_{4s}\sin\theta_\ell)\sin\theta_{\Lc}\cos\phi\big\}.
 \label{eq:baryondist}
\end{align}
\end{widetext}
Here $\dd\Gamma_B/\dd q^2$ is the dilepton-mass spectrum, and the normalization implies $2K_{1ss}+K_{1cc}=1$.
The integrated moment is the rate-weighted average
\begin{equation}
 \avg{K_{3s}}=
 \frac{\int\dd q^2\,(\dd\Gamma_B/\dd q^2)K_{3s}(q^2)}
      {\int\dd q^2\,(\dd\Gamma_B/\dd q^2)}.
\end{equation}
We use the unnormalized transversity amplitudes $A_{\lambda j}$ and $A^T_{\lambda j}$ of Appendix~\ref{app:baryon-amplitudes} throughout. Their bilinears determine the unnormalized coefficients ${\cal K}_i$, related to the coefficients in Eq.~\eqref{eq:baryondist} by
\begin{equation}
 K_i(q^2)=\frac{{\cal K}_i(q^2)}{\dd\Gamma_B/\dd q^2},
 \qquad \frac{\dd\Gamma_B}{\dd q^2}=2{\cal K}_{1ss}+{\cal K}_{1cc}.
 \label{eq:baryon-normalization}
\end{equation}
The form-factor amplitudes and the two coefficients in the denominator are given in Eqs.~\eqref{eq:lbamps}--\eqref{eq:baryonrate}. Thus $\avg{K_{3s}}=\int\dd q^2\,{\cal K}_{3s}/\int\dd q^2\,(\dd\Gamma_B/\dd q^2)$.

The baryonic sine coefficients identify the imaginary interferences used in our comparison. In the normalization of Eq.~\eqref{eq:baryondist}, their expressions follow from Ref.~\cite{Boer:2019zmp}.
The parameter $\alpha_{\Lc}$ is the parity-violating decay asymmetry of $\Lc\to\Lambda\pi$: it quantifies the forward--backward preference of the daughter $\Lambda$ relative to the parent spin.  For a $\Lc$ polarization vector $\bm P_{\Lc}$ and the unit vector $\widehat{\bm p}_\Lambda$ along the $\Lambda$ momentum in the $\Lc$ rest frame, $\dd\Gamma/\dd\Omega\propto1+\alpha_{\Lc}\bm P_{\Lc}\cdot\widehat{\bm p}_\Lambda$.
We use $\alpha_{\Lc}=-0.82\pm0.09$, the average constructed in Ref.~\cite{Boer:2019zmp} including the BESIII measurement ~\cite{BESIII:2019odb}.

Recall that $r_\ell=m_\ell/\sqrt{q^2}$.  Writing $(\perp\leftrightarrow\parallel)$ for exchange of all transversity labels in the immediately preceding bracket, the full mass-dependent numerator is
\begin{widetext}
\begin{align}
 {\cal K}_{3s}={}&-\frac{\alpha_{\Lc}}{\sqrt2}
 \im\!\left[A_{\perp0}A_{\parallel1}^*
       +r_\ell^2A_{\perp1}A_{\perp t}^*-(\perp\leftrightarrow\parallel)\right]\nonumber\\
 &+\sqrt2\alpha_{\Lc}r_\ell
 \im\!\left[A^T_{\perp0}A_{\parallel1}^*
       +A^T_{\perp1}A_{\parallel0}^*
       +A^T_{\perp1}A_{\perp t}^*-(\perp\leftrightarrow\parallel)\right]\nonumber\\
 &-2\sqrt2\alpha_{\Lc}r_\ell^2
\im\!\left[A^T_{\perp0}A^{T*}_{\parallel1}
-(\perp\leftrightarrow\parallel)\right].
 \label{eq:k3s}
\end{align}
\end{widetext}
This includes the vector--scalar and axial-vector--pseudoscalar terms whose presence was emphasized in Ref.~\cite{Boer:2019zmp}.
For real form factors, substitution of the amplitudes reduces this numerator to the following polynomial in total Wilson coefficients:
\begin{align}
 {\cal K}_{3s}={}&G_{VA}\im(\cV\cA^*)+
 G_{VS}\im(\cV\cS^*)\nonumber\\
 &+G_{AP}\im(\cA\cP^*)\nonumber\\
 &+G_{VT}\im(\cV\cT^*)+
 G_{AT}\im(\cA\cT^*)\nonumber\\
 &+G_{ST}\im(\cS\cT^*)+
 G_{PT}\im(\cP\cT^*).
 \label{eq:kmap}
\end{align}
Equations~\eqref{eq:lbamps} and \eqref{eq:lbtensor} specify all amplitudes needed to obtain the seven real coefficient functions.

The numerator of the other sine coefficient is
\begin{align}
 {\cal K}_{3sc}&={}-\frac{\alpha_{\Lc}}{\sqrt2}(1-r_\ell^2)
 \im\!\left[A_{\perp0}A_{\perp1}^*-A_{\parallel0}A_{\parallel1}^*\right]\nonumber\\
 &\,+2\sqrt2\alpha_{\Lc}(1-r_\ell^2)
 \im\!\left[A^T_{\perp0}A^{T*}_{\perp1}
 -A^T_{\parallel0}A^{T*}_{\parallel1}\right].\label{eq:k3sc}
\end{align}
Each product contains the same Wilson coefficient and its conjugate. Consequently,
\begin{equation}
 K_{3sc}=0 .
 \label{eq:k3sczero}
\end{equation}
The labels $\perp$ and $\parallel$ denote the two transversality sectors. The indices $j=1$ and $j=0$ denote transverse and longitudinal virtual-current polarizations, and $t$ denotes the timelike amplitude.  The superscript $T$ identifies a tensor amplitude.  For $j=0,1$, their short-distance content is $A_{\perp j}\propto\cV$, $A_{\parallel j}\propto\cA$, the timelike amplitudes additionally contain $\cS$ or $\cP$, and $A^T_{\lambda j}\propto\cT$ (see Appendix~\ref{app:baryon-amplitudes}).  Substitution into Eq.~\eqref{eq:k3s} therefore produces precisely the imaginary bilinears in Eq.~\eqref{eq:kmap}.  By contrast, every term in ${\cal K}_{3sc}$ contains one coefficient times its own complex conjugate, so its imaginary part vanishes.  For real form factors this result holds for arbitrary complex coefficients in the stated operator basis.

\subsection{Comparison of the two channels}
\label{sec:complementarity}
The coefficient maps in Eqs.~\eqref{eq:imap} and \eqref{eq:kmap} give the weak-phase combinations measured by the selected moments.
Table~\ref{tab:map} summarizes their interference terms, including the explicit lepton-mass factors. In particular, the baryonic coefficient contains vector--scalar, axial-vector--pseudoscalar, vector--tensor, and scalar--tensor terms absent from the three mesonic coefficients.
The mesonic coefficient also retains a vector--pseudoscalar term absent from this baryonic coefficient.
The scalar difference follows from $\langle D^*|\bar cb|\bar B\rangle=0$. The scalar matrix element in $\Lb\to\Lc$ is nonzero. The vector--scalar interference carries a factor $m_\ell/\sqrt{q^2}$, motivating the tau-channel analysis below.

\begin{table*}[t]
\caption{Imaginary-interference support of the selected unnormalized coefficients for real form factors. 
Each entry denotes the leading explicit power of $r_\ell=m_\ell/\sqrt{q^2}$ multiplying the corresponding bilinear. The entry ``1'' means no explicit helicity suppression. 
Kinematic zeros at endpoints and accidental cancellations are omitted.}
\label{tab:map}
\begin{ruledtabular}
\begin{tabular}{lcccc}
Imaginary bilinear & $I_7$ & $I_8$ & $I_9$ & ${\cal K}_{3s}$\\
\hline
$\im(\cV\cA^*)$ & $r_\ell^2$ & 1 & 1 & 1\\
$\im(\cV\cS^*)$ & 0 & 0 & 0 & $r_\ell$\\
$\im(\cV\cP^*)$ & $r_\ell$ & 0 & 0 & 0\\
$\im(\cV\cT^*)$ & 0 & 0 & 0 & $r_\ell$\\
$\im(\cA\cS^*)$ & 0 & 0 & 0 & 0\\
$\im(\cA\cP^*)$ & 0 & 0 & 0 & $r_\ell$\\
$\im(\cA\cT^*)$ & $r_\ell$ & 0 & 0 & $r_\ell$\\
$\im(\cS\cP^*)$ & 0 & 0 & 0 & 0\\
$\im(\cS\cT^*)$ & 0 & 0 & 0 & 1\\
$\im(\cP\cT^*)$ & 1 & 0 & 0 & 1
\end{tabular}
\end{ruledtabular}
\end{table*}

\section{Numerical analysis}
\label{sec:numerics}
\begin{figure*}[t]
\centering
\includegraphics[width=.94\textwidth]{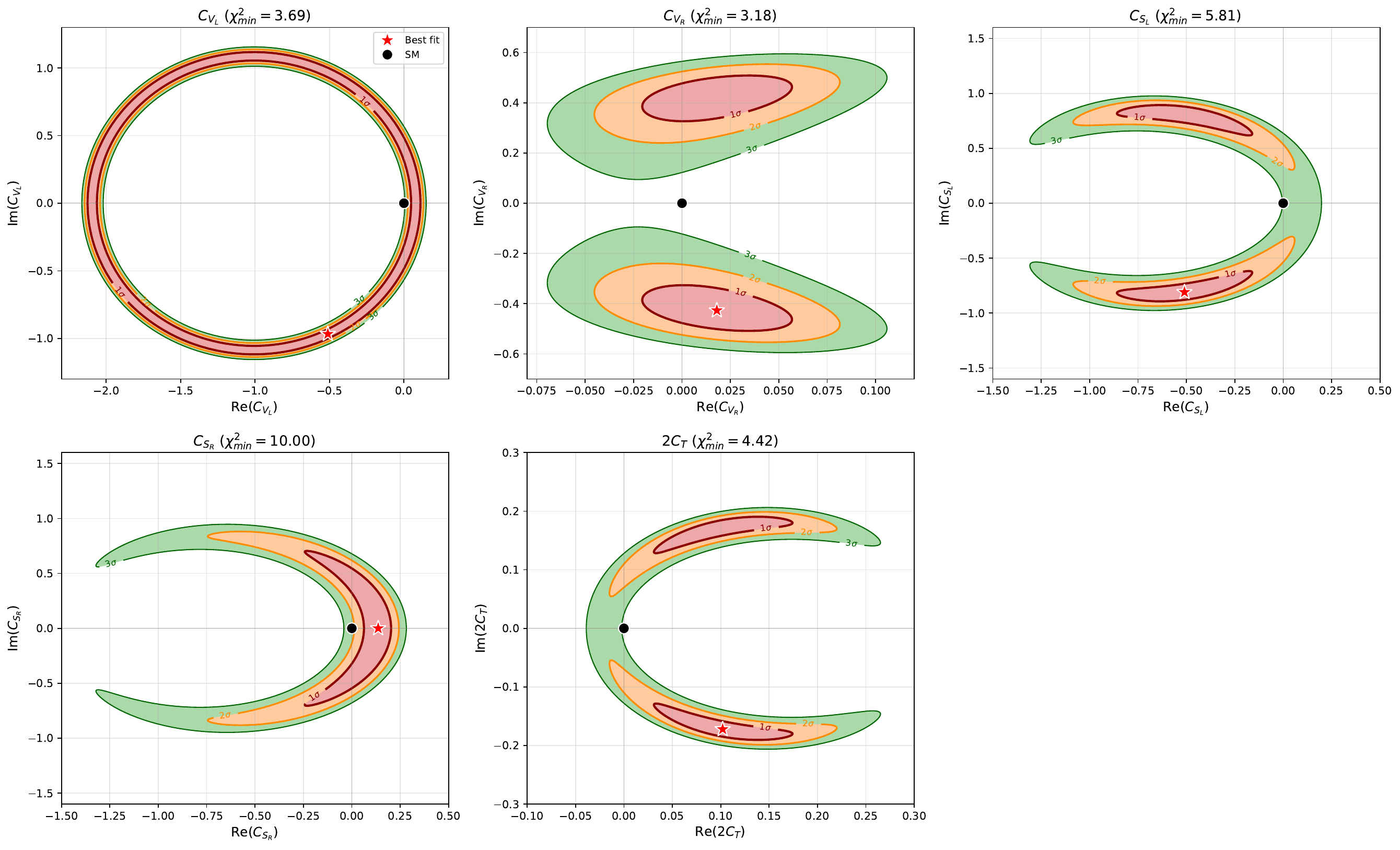}
\caption{$C\!P$-even fit contours for the chiral-basis coefficients $C_{V_L},C_{V_R},C_{S_L},C_{S_R},2C_T$. The tensor panel shows $\re(2C_T)$ versus $\im(2C_T)$. Each panel varies one complex NP coefficient at $\mu_b=4.8$ GeV, with the other NP coefficients set to zero. The nested regions correspond to the $1\sigma$, $2\sigma$, and $3\sigma$ confidence levels for two fitted parameters, defined by $\Delta\chi^2=2.30$, $6.18$, and $11.83$, respectively.}
\label{fig:fit-chiral}
\end{figure*}

\begin{figure*}[t]
\centering
\includegraphics[width=.94\textwidth]{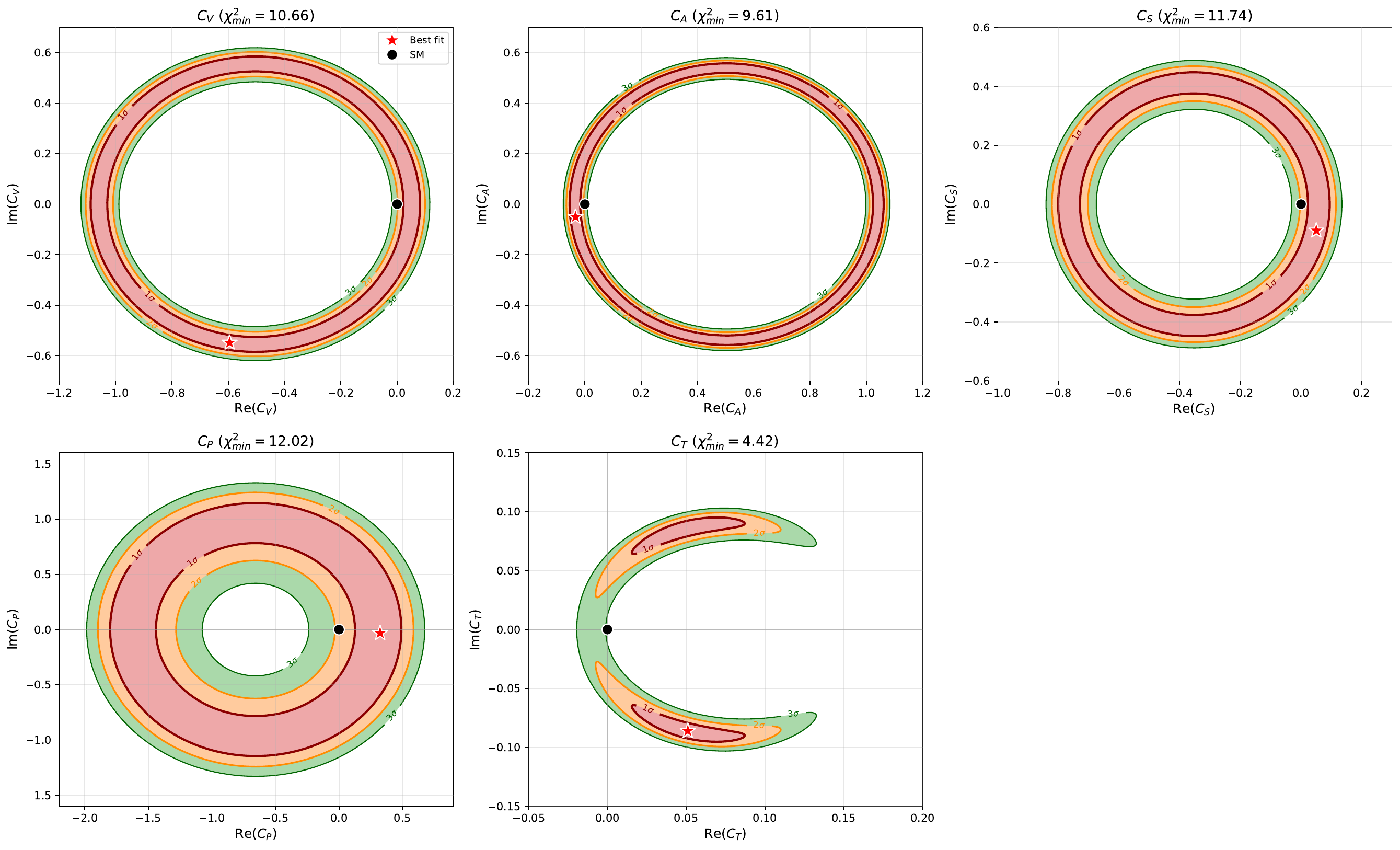}
\caption{$C\!P$-even fit contours in the parity basis for $C_V,C_A,C_S,C_P,C_T$, with $C_{V,A,S,P}$ defined in Eq.~\eqref{eq:parity-np} and $C_T$ normalized as in Eq.~\eqref{eq:ham}. The tensor panel shows $\re C_T$ versus $\im C_T$. The inputs and confidence levels are the same as in Fig.~\ref{fig:fit-chiral}. Each panel is a separate single-coefficient hypothesis.}
\label{fig:fit-parity}
\end{figure*}

{
Fits to $b\to c\tau\bar\nu_\tau$ rates and polarization observables provide complementary constraints on the Lorentz structure of the charged current~\cite{Tanaka:2012nw,Sakaki:2013bfa,Murgui:2019czp,Hu:2018veh,Yan:2019hpm,Wang:2019trs,Kumbhakar:2020jdz,Iguro:2024hyk}. Correlations between mesonic and baryonic modes have also been studied in effective-theory and model-specific analyses~\cite{Datta:2017aue,Hu:2020axt,Duan:2024ayo}. We use the measured $C\!P$-even observables to constrain one complex Wilson coefficient at a time, first in the chiral basis and then in the parity basis. In each scenario, its real and imaginary parts are varied while all other NP coefficients in the chosen basis are set to zero. NP is restricted to the tau channel.

For the parity-basis (V/A/S/P/T) fits, $C_T$ has the normalization specified in Sec.~\ref{sec:eft}, and the remaining NP coefficients are
\begin{align}
 C_V&=(C_{V_L}+C_{V_R})/2,&
 C_A&=(C_{V_R}-C_{V_L})/2,\nonumber\\
 C_S&=(C_{S_L}+C_{S_R})/2,&
 C_P&=(C_{S_R}-C_{S_L})/2. \label{eq:parity-np}
\end{align}
Here $C_V,C_A,C_S,C_P,C_T$ multiply the corresponding quark currents with the common prefactor $4G_FV_{cb}/\sqrt2$ and the chiral leptonic currents kept unchanged. They denote NP coefficients, whereas $\cV,\cA,\cS,\cP,\cT$ denote the total amplitude coefficients of Eq.~\eqref{eq:parity}.
At a common scale and with the SM normalization in Eq.~\eqref{eq:ham}, these are related to the total amplitude coefficients by
\begin{align}
 \cV&=1+2C_V,& \cA&=1-2C_A,\nonumber\\
 \cS&=2C_S,& \cP&=2C_P,\nonumber\\
 \cT&=2C_T.&&
\end{align}
The minus sign in $\cA=1-2C_A$ follows from writing the axial quark current as $-\cA\,\bar c\gamma^\mu\gamma_5b$. The vector and scalar single-coefficient fits in the two bases describe different hypotheses: a pure $C_S=s$ direction has $C_{S_L}=C_{S_R}=s$, whereas a single nonzero $C_{S_L}$ contains both scalar and pseudoscalar currents. The tensor fits describe the same interaction, displayed in the coordinate $2C_T$ for the chiral basis and $C_T$ for the parity basis.

We use the benchmark input set in Table~\ref{tab:inputs}. The rate predictions are evaluated with \texttt{EOS}~\cite{EOSAuthors:2021xpv} using the shared experimental and theory inputs of the reference \texttt{flavio} likelihood~\cite{Straub:2018kue}. The NP coefficients are specified directly at $\mu_b=4.8$ GeV. The fit adds the low-scale SM vector coefficient $C_{V_L}^{\rm SM}=1.00709$ to the NP contribution. The experimental constraints are treated as independent Gaussians, with statistical and systematic errors combined in quadrature. The covariance of the supported theory nuisance parameters is evaluated at the SM point and held fixed across the scan and added to the experimental covariance. With $\bm O_{\rm exp}$ and $\bm O_{\rm th}(C)$ denoting the input and prediction vectors, we use
\begin{align}
 \chi^2(C)&=\bm\Delta(C)^T
 (\Sigma_{\rm exp}+\Sigma_{\rm th}^{\rm SM})^{-1}\bm\Delta(C),\nonumber\\
 \bm\Delta(C)&=\bm O_{\rm th}(C)-\bm O_{\rm exp}.
\end{align}

\begin{table}[tb]
\caption{Benchmark $C\!P$-even inputs used for the displayed fit contours~\cite{LHCb:2023ssl,LHCb:2026dev,Bernlochner:2017jka,Bordone:2019vic,Harrison:2020gvo}.  }
\label{tab:inputs}
\begin{ruledtabular}
\begin{tabular}{lc}
Observable & Input\\ \hline
$R(D)$ & $0.359\pm0.024$\\
$R(D^*)$ & $0.280\pm0.011$\\
$P_\tau(D^*)$ & $-0.298\pm0.481$\\
$F_L(D^*)$ & $0.41\pm0.06\pm0.03$\\
$R(J/\psi)$ & $0.51\pm0.12\pm0.08$
\end{tabular}
\end{ruledtabular}
\end{table}

The $R(D^{(*)})$, $F_L(D^*)$, and $P_\tau(D^*)$ predictions share the correlated 23-parameter HQET form-factor input~\cite{Bernlochner:2017jka,Bordone:2019vic}, including its contribution to their theory covariance. The $R(J/\psi)$ prediction uses fixed HPQCD vector and axial-vector form factors~\cite{Harrison:2020gvo} and vanishing tensor form factors. The fixed $B_c\to J/\psi$ form-factor errors are therefore not included in the fit covariance. These assumptions define the benchmark contours. The angular predictions below use the lattice inputs described in Sec.~\ref{sec:ff} and the unit-SM normalization of Eq.~\eqref{eq:ham}.

Fig.~\ref{fig:fit-chiral} and Fig.~\ref{fig:fit-parity} show the two sets of fits. The contours are defined by $\Delta\chi^2=\chi^2-\chi^2_{\min}$, with thresholds 2.30, 6.18, and 11.83 for nominal $68.3\%$, $95.4\%$, and $99.7\%$ regions in two fit parameters. For $C_{V_L}$, $C_{V_R}$, $C_{S_L}$, $C_{S_R}$, and $2C_T$, the minimum $\chi^2$ values are 3.69, 3.18, 5.81, 10.00, and 4.42, respectively. For $C_V$, $C_A$, $C_S$, $C_P$, and $C_T$, they are 10.66, 9.61, 11.74, 12.02, and 4.42. The common SM value is $\chi^2_{\rm SM}=17.34$. Each confidence region is defined by the increase in $\chi^2$ from the minimum of its own single-coefficient scenario.
}


We next evaluate the normalized integrated moments $\avg{I_7}$, $\avg{I_8}$, $\avg{I_9}$, and $\avg{K_{3s}}$ over the full physical tau phase space, normalizing each integrated angular numerator to the corresponding integrated rate. The following numbers reproduce the supplied angular results for these two sets of benchmark coefficients. 

Table~\ref{tab:best-moments} collects the supplied benchmark values and their form-factor errors for both bases. We display representative best-fit grid points with negative imaginary part when present. The $C_{S_R}$ minimum lies on the real axis. For complex points, their complex conjugates give identical $C\!P$-even predictions and reverse all four moments. The $C_{V_L}$ and parity-basis $V$, $A$, and $S$ fits also admit continuous degeneracies, so their listed coordinates specify representative points.

\begin{table*}[t]

\caption{Supplied benchmark values of the normalized integrated $C\!P$-odd angular moments, in units of $10^{-2}$. The upper and lower blocks use the chiral and parity bases, respectively; their tensor coordinates are $2C_T$ and $C_T$. Errors are one-standard-deviation form-factor uncertainties at fixed Wilson coefficients. The baryonic errors use the nominal lattice covariance and exclude the independent $\alpha_{\Lc}$ uncertainty. NP coefficients are quoted at $4.8$ GeV and the central analyzer is $\alpha_{\Lc}=-0.82$. 
}
\label{tab:best-moments}
\small
\setlength{\tabcolsep}{4pt}
\begin{ruledtabular}
\begin{tabular}{lccccc}
Coefficient & Grid minimum & $\avg{I_7}$ & $\avg{I_8}$ & $\avg{I_9}$ & $\avg{K_{3s}}$\\ \hline
$C_{V_L}$ & $-0.5112-0.9672i$ & $0.00$ & $0.00$ & $0.00$ & $0.00$ \\
$C_{V_R}$ & $+0.0180-0.4270i$ & $+3.00\pm0.32$ & $-3.04\pm0.23$ & $-5.53\pm0.45$ & $-10.38\pm0.08$ \\
$C_{S_L}$ & $-0.5100-0.8096i$ & $+2.23\pm0.22$ & $0.00$ & $0.00$ & $-4.26\pm0.05$ \\
$C_{S_R}$ & $+0.1360$ & $0.00$ & $0.00$ & $0.00$ & $0.00$ \\
$2C_T$ & $+0.1020-0.1722i$ & $+2.25\pm0.62$ & $0.00$ & $0.00$ & $-3.25\pm0.07$ \\
\hline
$C_V$ & $-0.5952-0.5488i$ & $+4.38\pm0.47$ & $-4.45\pm0.33$ & $-8.08\pm0.65$ & $-14.48\pm0.11$ \\
$C_A$ & $-0.0334-0.0504i$ & $+0.36\pm0.04$ & $-0.36\pm0.03$ & $-0.66\pm0.06$ & $-1.30\pm0.01$ \\
$C_S$ & $+0.0504-0.0900i$ & $0.00$ & $0.00$ & $0.00$ & $-0.58\pm0.01$ \\
$C_P$ & $+0.3234-0.0320i$ & $-0.18\pm0.02$ & $0.00$ & $0.00$ & $+0.16\pm0.00$ \\
$C_T$ & $+0.0510-0.0861i$ & $+2.25\pm0.62$ & $0.00$ & $0.00$ & $-3.25\pm0.07$ \\
\end{tabular}
\end{ruledtabular}
\end{table*}

A common left-handed vector rescaling gives zero for every moment because it introduces no relative weak phase. In the scalar, pseudoscalar, and tensor single-coefficient benchmarks, $I_8$ and $I_9$ also vanish, in agreement with Eq.~\eqref{eq:imap}. 

The ranges in Tables~\ref{tab:ranges-chiral} and \ref{tab:ranges-operator} are obtained by mapping the two-dimensional Wilson-coefficient regions in Figs.~\ref{fig:fit-chiral} and \ref{fig:fit-parity} to the angular observables at central form factors. The two disjoint intervals arise from the separated positive- and negative-imaginary-part intervals of the allowed Wilson coefficients. Uncertainties from form factors and other theoretical inputs are not taken into account at this stage. The fixed-coefficient errors in Table~\ref{tab:best-moments} are small for the baryonic benchmarks and modest for most mesonic benchmarks, although the tensor contribution to $\avg{I_7}$ has a larger relative uncertainty.

\begin{table*}[t]
\caption{Chiral-basis ranges of the normalized integrated moments, in units of $10^{-2}$, obtained by mapping the two-parameter Wilson-coefficient regions at central hadronic inputs. Here $\pm[a,b]$ denotes $[-b,-a]\cup[a,b]$. The quoted levels refer to the original two-dimensional fit regions. }
\label{tab:ranges-chiral}
\begin{ruledtabular}
\begin{tabular}{lccccc}
Varied coefficient & Region & $\avg{I_7}$ & $\avg{I_8}$ & $\avg{I_9}$ & $\avg{K_{3s}}$\\ \hline
$C_{V_L}$ & $68.3\%$ & $0.00$ & $0.00$ & $0.00$ & $0.00$ \\
 & $95.4\%$ & $0.00$ & $0.00$ & $0.00$ & $0.00$ \\
 & $99.7\%$ & $0.00$ & $0.00$ & $0.00$ & $0.00$ \\
\noalign{\smallskip}
$C_{V_R}$ & $68.3\%$ & $\pm[2.38,3.47]$ & $\pm[2.41,3.52]$ & $\pm[4.38,6.39]$ & $\pm[8.39,11.73]$ \\
 & $95.4\%$ & $\pm[1.79,3.72]$ & $\pm[1.82,3.77]$ & $\pm[3.30,6.86]$ & $\pm[6.40,12.37]$ \\
 & $99.7\%$ & $\pm[0.73,3.94]$ & $\pm[0.74,4.00]$ & $\pm[1.34,7.26]$ & $\pm[2.63,12.89]$ \\
\noalign{\smallskip}
$C_{S_L}$ & $68.3\%$ & $\pm[1.78,2.43]$ & $0.00$ & $0.00$ & $\pm[3.45,4.47]$ \\
 & $95.4\%$ & $\pm[0.90,2.53]$ & $0.00$ & $0.00$ & $\pm[1.76,4.57]$ \\
 & $99.7\%$ & $[-2.64,+2.64]$ & $0.00$ & $0.00$ & $[-4.65,+4.65]$ \\
\noalign{\smallskip}
$C_{S_R}$ & $68.3\%$ & $[-2.10,+2.10]$ & $0.00$ & $0.00$ & $[-0.45,+0.45]$ \\
 & $95.4\%$ & $[-2.66,+2.66]$ & $0.00$ & $0.00$ & $[-0.55,+0.55]$ \\
 & $99.7\%$ & $[-2.86,+2.86]$ & $0.00$ & $0.00$ & $[-0.57,+0.57]$ \\
\noalign{\smallskip}
$2C_T$ & $68.3\%$ & $\pm[1.68,2.36]$ & $0.00$ & $0.00$ & $\pm[2.47,3.45]$ \\
 & $95.4\%$ & $\pm[0.71,2.36]$ & $0.00$ & $0.00$ & $\pm[1.05,3.50]$ \\
 & $99.7\%$ & $[-2.36,+2.36]$ & $0.00$ & $0.00$ & $[-3.53,+3.53]$ \\
\end{tabular}
\end{ruledtabular}
\end{table*}

\begin{table*}[t]
\caption{Parity-basis ranges in units of $10^{-2}$, with the same confidence levels, interval notation, and central-hadronic-input prescription as Table~\ref{tab:ranges-chiral}.}
\label{tab:ranges-operator}
\begin{ruledtabular}
\begin{tabular}{lccccc}
Varied coefficient & Region & $\avg{I_7}$ & $\avg{I_8}$ & $\avg{I_9}$ & $\avg{K_{3s}}$\\ \hline
$C_V$ & $68.3\%$ & $[-4.63,+4.63]$ & $[-4.70,+4.70]$ & $[-8.54,+8.54]$ & $[-14.87,+14.87]$ \\
 & $95.4\%$ & $[-4.76,+4.76]$ & $[-4.83,+4.83]$ & $[-8.77,+8.77]$ & $[-14.96,+14.96]$ \\
 & $99.7\%$ & $[-4.87,+4.87]$ & $[-4.94,+4.94]$ & $[-8.97,+8.97]$ & $[-15.04,+15.04]$ \\
\noalign{\smallskip}
$C_A$ & $68.3\%$ & $[-3.90,+3.90]$ & $[-3.96,+3.96]$ & $[-7.19,+7.19]$ & $[-14.00,+14.00]$ \\
 & $95.4\%$ & $[-3.99,+3.99]$ & $[-4.04,+4.04]$ & $[-7.35,+7.35]$ & $[-14.13,+14.13]$ \\
 & $99.7\%$ & $[-4.08,+4.08]$ & $[-4.14,+4.14]$ & $[-7.52,+7.52]$ & $[-14.26,+14.26]$ \\
\noalign{\smallskip}
$C_S$ & $68.3\%$ & $0.00$ & $0.00$ & $0.00$ & $[-2.67,+2.67]$ \\
 & $95.4\%$ & $0.00$ & $0.00$ & $0.00$ & $[-2.72,+2.72]$ \\
 & $99.7\%$ & $0.00$ & $0.00$ & $0.00$ & $[-2.77,+2.77]$ \\
\noalign{\smallskip}
$C_P$ & $68.3\%$ & $[-6.07,+6.07]$ & $0.00$ & $0.00$ & $[-5.63,+5.63]$ \\
 & $95.4\%$ & $[-6.40,+6.40]$ & $0.00$ & $0.00$ & $[-5.97,+5.97]$ \\
 & $99.7\%$ & $[-6.68,+6.68]$ & $0.00$ & $0.00$ & $[-6.25,+6.25]$ \\
\noalign{\smallskip}
$C_T$ & $68.3\%$ & $\pm[1.68,2.36]$ & $0.00$ & $0.00$ & $\pm[2.47,3.45]$ \\
 & $95.4\%$ & $\pm[0.71,2.36]$ & $0.00$ & $0.00$ & $\pm[1.06,3.50]$ \\
 & $99.7\%$ & $[-2.36,+2.36]$ & $0.00$ & $0.00$ & $[-3.53,+3.53]$ \\
\end{tabular}
\end{ruledtabular}
\end{table*}


The first implication is that the $C\!P$-even fit and the $C\!P$-odd angular moments probe different parts of the same amplitude products. Rates and polarization observables depend on modulus squares and real interference terms, while the angular numerators in Eqs.~\eqref{eq:imap} and \eqref{eq:kmap} are linear in imaginary bilinears $\im(\mathcal C_i\mathcal C_j^*)$. With real hadronic functions, complex conjugation leaves the former unchanged and reverses the latter. Thus a nonzero angular measurement can distinguish conjugate solutions that the rate fit cannot separate. For instance, the $C_{V_R}$ region maps at $68.3\%$ to $\avg{I_9}\in\pm[0.0438,0.0639]$ and $\avg{K_{3s}}\in\pm[0.0839,0.1173]$. 

The second implication is the scalar sensitivity unique to the baryonic channel in this comparison. The $\bar B\to D^*$ scalar matrix element vanishes, so $I_{7,8,9}$ are independent of $\cS$. The baryonic coefficient $K_{3s}$ instead contains $\im(\cV\cS^*)$ and, when a tensor coupling is also present, $\im(\cS\cT^*)$. In the pure-scalar benchmark, all three mesonic moments vanish throughout the allowed region, while $\avg{K_{3s}}$ spans $[-0.0267,0.0267]$ at $68.3\%$ and $[-0.0272,0.0272]$ at $95.4\%$. This comparison concerns the scalar sum $C_{S_L}+C_{S_R}$. The pseudoscalar difference remains accessible through $I_7$, and $R(D)$ already constrains the scalar coupling through a $C\!P$-even observable. A baryonic $C\!P$-odd angular moment therefore supplies scalar-phase information absent from the selected $D^*$ moments. The single-coefficient benchmarks illustrate this complementarity without establishing that all phases can be separated in a general multi-coefficient fit.

\section{Experimental prospects}
\label{sec:discussion}
We estimate the statistical reach of a signed angular count in an ideal signal sample with known production angles and uniform efficiency. Integrating Eq.~\eqref{eq:baryondist} over the polar angles gives
\begin{equation}
 \frac{\dd\Gamma_+(q^2)-\dd\Gamma_-(q^2)}{\dd\Gamma_+(q^2)+\dd\Gamma_-(q^2)}
 = \frac{3\pi}{8}K_{3s}(q^2).
 \label{eq:counting}
\end{equation}
Here $\dd\Gamma_\pm$ are the differential partial rates for $\sin\phi\gtrless0$, $\phi$ is the oriented angle defined in Sec.~\ref{sec:baryon}, and $K_{3s}$ is the normalized coefficient of $\sin\theta_\ell\sin\theta_{\Lc}\sin\phi$.

Integrating over the tau phase space yields
\begin{equation}
 A_{\rm LR}\equiv\frac{N_+-N_-}{N_++N_-}
 = \frac{3\pi}{8}\avg{K_{3s}}.
 \label{eq:alr}
\end{equation}
Here $A_{\rm LR}$ is the signed-yield asymmetry, $N_+$ and $N_-$ count signal events with positive and negative $\sin\phi$, and $\avg{K_{3s}}$ is the rate-weighted integrated moment. Comparing charge-conjugate samples in $C\!P$-mapped angular and daughter-analyzer conventions gives $A_{\rm LR}^{\rm CP}=(A_{\rm LR}-\overline A_{\rm LR}^{\,\rm CP})/2$, where $\overline A_{\rm LR}^{\,\rm CP}$ is the conjugate-sample asymmetry. For real hadronic functions and $C\!P$ conservation in the secondary decay, $A_{\rm LR}^{\rm CP}=A_{\rm LR}$.

Binomial counting then gives
\begin{equation}
 \sigma_A^2=\frac{1-A^2}{N},\qquad
 N\simeq\frac{Z^2(1-A^2)}{A^2}.
 \label{eq:statistics}
\end{equation}
Here $A$ is the assumed asymmetry, $\sigma_A$ its statistical standard deviation, $N$ the signal-event count, and $Z=|A|/\sigma_A$ the target Gaussian sensitivity. For $A=A_{\rm LR}$, $N=N_++N_-$. For $A=A_{\rm LR}^{\rm CP}$, $N$ is the combined count of two equally sized conjugate samples.
The largest $68.3\%$ vector and scalar moments in the parity basis correspond to $|A|\simeq17.5\%$ and $3.14\%$. Equation~\eqref{eq:statistics} then gives about $7.9\times10^2$ and $2.5\times10^4$ effective signal events for $Z=5$.

LHCb reconstructed $349\pm40$ $\Lb\to\Lc\tau\bar\nu$ signal decays in $3\,\mathrm{fb}^{-1}$ of Run~1 data, using $\Lc\to pK^-\pi^+$ and $\tau\to3\pi(\pi^0)\nu$, including charge-conjugate modes~\cite{LHCb:2022piu}. Using the central yield as a reference and holding the semitauonic branching fraction fixed, we write
\begin{equation}
 N_{\rm eff}\simeq349\,
 \frac{\mathcal L}{3\,\mathrm{fb}^{-1}}\,
 r_\sigma r_\epsilon.
 \label{eq:yield-scaling}
\end{equation}
Here $N_{\rm eff}$ is the combined effective signal count to use in Eq.~\eqref{eq:statistics}, $\mathcal L$ the integrated luminosity, $r_\sigma=\sigma_{\Lb}/\sigma_{\Lb}^{\rm Run\,1}$ the production-cross-section ratio within the acceptance, and $r_\epsilon=\epsilon_{\rm eff}/\epsilon_{\rm Run\,1}$ the effective efficiency relative to the Run~1 yield reference. Both efficiencies include daughter branching fractions and reconstruction and selection efficiencies. The effective efficiency $\epsilon_{\rm eff}$ additionally accounts for loss of angular information and the statistical cost of background subtraction. In particular, it contains the tau-cascade response~\cite{Ligeti:2016npd,Bhattacharya:2020lfm,Hu:2020axt,Ciezarek:2016lqu} and the change from $\Lc\to pK^-\pi^+$ to the $\Lc\to\Lambda\pi^+$ analyzer. Thus $r_\epsilon=1$ means the same effective yield per produced semitauonic decay as the Run~1 reference.

For the $300\,\mathrm{fb}^{-1}$ luminosity benchmark of LHCb Upgrade~II~\cite{Vagnoni:2025qfv}, taking $r_\sigma=1$ and $r_\epsilon=1$ gives $N_{\rm eff}\simeq3.49\times10^4$. If the pure-scalar asymmetry reaches $|A|=3.14\%$, this corresponds to an ideal statistical sensitivity of $5.9\sigma$. More generally, at this luminosity and production cross section, a $5\sigma$ sensitivity requires $r_\epsilon\gtrsim0.73$, or an overall effective efficiency at least $73\%$ of the Run~1 yield-reference efficiency. This condition gives a quantitative target for an angular analysis. Its feasibility depends on the channel-specific tau and $\Lc$ reconstruction and angular response.

For $\bar B\to D^*(\to D\pi)\tau^-\bar\nu_\tau$, the same signed-yield construction extracts $\avg{I_{7,8,9}}$. Define $A_7$, $A_8$, and $A_9$ using the ratio in Eq.~\eqref{eq:alr}, with $N_\pm$ selected by the sign of $\sin\chi\cos\theta_D$, $\sin\chi\cos\theta_D\cos\theta_\ell$, and $\sin2\chi$, respectively. Integrating Eq.~\eqref{eq:mesondist} over the full tau phase space gives
\begin{align}
 A_7&=\frac34\avg{I_7},&
 A_8&=\frac{2}{\pi}\avg{I_8},&
 A_9&=\frac{2}{\pi}\avg{I_9}.
 \label{eq:meson-counting}
\end{align}
The same $C\!P$-mapped conjugate subtraction and statistical estimate in Eq.~\eqref{eq:statistics} apply to each asymmetry. At the largest $68.3\%$ values in Table~\ref{tab:ranges-operator}, the vector benchmark gives $|A_9|\simeq5.44\%$ and the tensor benchmark gives $|A_7|\simeq1.77\%$, requiring approximately $8.4\times10^3$ and $8.0\times10^4$ effective signal events, respectively, for $Z=5$.

For Belle~II, Ref.~\cite{Hill:2019zja} projects about $7000$ reconstructed signal events with hadronic tagging and $\tau\to3\pi(\pi^0)\nu$ at $50\,\mathrm{ab}^{-1}$. Using this yield in Eq.~\eqref{eq:statistics} gives ideal statistical sensitivities of $4.6\sigma$ and $1.5\sigma$ for the vector $A_9$ and tensor $A_7$ benchmarks, respectively. Scaling linearly with luminosity, the corresponding ideal $5\sigma$ thresholds are approximately $60\,\mathrm{ab}^{-1}$ and $570\,\mathrm{ab}^{-1}$, before accounting for angular-information loss and background subtraction.

\section{Conclusions}

We have compared the angular distributions of $\bar B\to D^*(\to D\pi)\ell^-\bar\nu_\ell$ and $\Lambda_b\to\Lambda_c(\to\Lambda\pi)\ell^-\bar\nu_\ell$ in a common effective Hamiltonian with left-handed neutrinos. Using correlated lattice-QCD form factors, with a weak-unitarity-constrained BGL description of the mesonic input, we identify the imaginary Wilson-coefficient products probed by each channel. The numerical analysis focuses on the tau mode and uses $C\!P$-even data to constrain one complex coefficient at a time in the chiral and parity bases. The same formalism can be used to study $C\!P$-odd angular observables in electron and muon modes, with numerical predictions requiring the corresponding experimental constraints and lepton-mass dependence.

The two angular distributions give complementary access to weak phases. The baryonic coefficient $K_{3s}$ probes the $VS$, $AP$, $VT$, and $ST$ interferences absent from the mesonic moments $I_{7,8,9}$, while $I_7$ provides sensitivity to the $VP$ interference. The scalar sum $C_{S_L}+C_{S_R}$ illustrates this distinction particularly clearly: it does not enter the $\bar B\to D^*$ amplitude but can generate a baryonic $C\!P$-odd angular moment. In the pure-scalar benchmark, the allowed region gives $|\avg{K_{3s}}|\lesssim2.67\times10^{-2}$ at $68.3\%$, whereas all three mesonic moments vanish. A measurement of $K_{3s}$ would therefore add scalar-phase information to the constraints already supplied by $R(D)$ and the $D^*$ angular distribution.

With real hadronic functions, the $C\!P$-odd angular moments change sign between complex-conjugate solutions that remain degenerate in the $C\!P$-even fit. Their combined measurement can distinguish these solutions and test the interference pattern of a proposed NP interaction. Extracting these moments experimentally requires the measurable distributions of the visible final-state particles in the full cascade decay.
In the present study, the Wilson-coefficient fits and the mapping to angular-observable ranges are performed separately, although the mesonic observables in both steps depend on common form factors. A complete NP analysis should determine the coefficients and angular ranges jointly, retaining these shared hadronic correlations. Our benchmarks illustrate the additional sensitivity of baryonic observables. A joint treatment is left for future work.

\begin{acknowledgments}
This work is supported by the National Natural Science Foundation of China under Grant Nos.~12522506, 12375086, 12447154, 12135006, 12635005, 12575099, 12335003, 12247101 and by the the Fundamental Research Funds for the Central Universities under No. lzujbky-2023-stlt01, lzujbky-2024-oy02, lzujbky-2025-jdzx07 and lzujbky-2025-eyt01, the Scientific Research Innovation Capability Support Project for Young Faculty under Grant No. ZYGXQNJSKYCXNLZCXMP2, the Natural Science Foundation of Gansu Province (No.25JRRA799), and the ‘111 Center’ under Grant No. B20063. We also acknowledge the High Energy Physics Base and its AI Club at the School of Physics, Huazhong University of Science and Technology for providing a collaborative discussion platform and computing/research resources.

\end{acknowledgments}

\appendix
\section{Helicity and transversity amplitude conventions}
\label{app:amplitudes}

This appendix collects the amplitude conventions required by both
angular distributions. All tensor amplitudes below use $\cT=2C_T$ as defined in Eq.~\eqref{eq:parity}.

\subsection{Mesonic helicity amplitudes}\label{app:meson-amplitudes}
In the angular convention of Sec.~\ref{sec:meson}, the subscripts $\pm,0,t$ denote transverse, longitudinal, and timelike virtual-current helicities. We use $\lambda=(m_B^2+m_{D^*}^2-q^2)^2-4m_B^2m_{D^*}^2$.  
The amplitudes are
\begin{align}
 H_\pm&=(m_B+m_{D^*})A_1\cA\mp\frac{\sqrt\lambda}{m_B+m_{D^*}}V\cV,\nonumber\\
 H_0&=-\frac{8m_B m_{D^*}}{\sqrt{q^2}}A_{12}\cA,\qquad
 H_t=-\sqrt{\frac{\lambda}{q^2}}A_0\cA,\nonumber\\
 H_P&=-\frac{\sqrt\lambda}{m_b+m_c}A_0\cP,\nonumber\\
 H_{T,\pm}&=\frac{2}{\sqrt{q^2}}
       [\pm\sqrt\lambda T_1+(m_B^2-m_{D^*}^2)T_2]\cT,\nonumber\\
 H_{T,0}&=-\frac{8m_B m_{D^*}}{m_B+m_{D^*}}T_{23}\cT.
 \label{eq:mesonamps}
\end{align}
The lepton-mass-dressed combinations entering
Eq.~\eqref{eq:i9} are
\begin{align}
 H_\lambda^-&=H_\lambda-2r_\ell H_{T,\lambda},&
 H_\lambda^+&=H_\lambda-\frac{2}{r_\ell}H_{T,\lambda},\nonumber\\
 \widetilde H_t&=H_t+\frac{H_P}{r_\ell},&
 r_\ell&=\frac{m_\ell}{\sqrt{q^2}}.\label{eq:mesondressed}
\end{align}
The superscripts $\pm$ label the two lepton-helicity combinations and are distinct from the helicity subscripts. An asterisk denotes complex conjugation, for example $H_0^{-*}\equiv(H_0^-)^*$. The relative signs of $H_0$, $H_t$, and the transverse amplitudes follow the angular definitions in Eq.~\eqref{eq:mesondist}.
The common phase-space and weak-normalization factor is
\begin{equation}
 {\cal N}_M(q^2)=\frac{G_F^2|V_{cb}|^2q^2\sqrt\lambda}{3\,2^7\pi^3m_B^3}
 (1-r_\ell^2)^2\,{\cal B}(D^*\to D\pi).
 \label{eq:mesonnorm}
\end{equation}
The differential rate corresponding to Eq.~\eqref{eq:i9} is
\begin{align}
 \frac{\dd\Gamma_M}{\dd q^2}=&{\cal N}_M\big[
 2(|H_0^-|^2+|H_+^-|^2+|H_-^-|^2)\nonumber\\
&+r_\ell^2(|H_0^+|^2+|H_+^+|^2+|H_-^+|^2
    +3|\widetilde H_t|^2)\big].\label{eq:mesonrate}
\end{align}
These expressions determine the integrated ratios $\avg{I_i}$ with the physical integration range $m_\ell^2\le q^2\le(m_B-m_{D^*})^2$.

\subsection{Baryonic transversity amplitudes and rate}\label{app:baryon-amplitudes}
Using the transversity labels of Sec.~\ref{sec:baryon} and $s_\pm=(m_{\Lb}\pm m_{\Lc})^2-q^2$, we choose a real common normalization $N_B$ with
\begin{equation}
 N_B^2=\frac{G_F^2|V_{cb}|^2q^2\sqrt{s_+s_-}}{3\,2^7\pi^3 m_{\Lb}^3} (1-r_\ell^2)^2\,{\cal B}(\Lc\to\Lambda\pi).
\end{equation}
These are the same unnormalized amplitudes used in Eqs.~\eqref{eq:k3s} and \eqref{eq:k3sc}. For $A_{\lambda j}$, $\lambda=\perp,\parallel$ and $j=0,1,t$; the tensor amplitudes have $j=0,1$. The six vector, axial-vector, and timelike amplitudes are
\begin{align}
A_{\perp1}&=-2N_B\sqrt{s_-}\,f_\perp\,\cV,\nonumber\\
A_{\parallel1}&=2N_B\sqrt{s_+}\,g_\perp\,\cA,\nonumber\\
A_{\perp0}&=\sqrt2N_B\sqrt{s_-}\frac{m_{\Lb}+m_{\Lc}}{\sqrt{q^2}}f_+\,\cV,\nonumber\\
A_{\parallel0}&=-\sqrt2N_B\sqrt{s_+}\frac{m_{\Lb}-m_{\Lc}}{\sqrt{q^2}}g_+\,\cA,\nonumber\\
A_{\perp t}&=\sqrt2N_B\sqrt{s_+}\frac{m_{\Lb}-m_{\Lc}}{\sqrt{q^2}}f_0\nonumber\\
&\quad\times\left(\cV+\frac{q^2\cS}{m_\ell(m_b-m_c)}\right),\nonumber\\
A_{\parallel t}&=-\sqrt2N_B\sqrt{s_-}\frac{m_{\Lb}+m_{\Lc}}{\sqrt{q^2}}g_0\nonumber\\
&\quad\times\left(\cA+\frac{q^2\cP}{m_\ell(m_b+m_c)}\right).
 \label{eq:lbamps}
\end{align}
The tensor amplitudes are
\begin{align}
A^T_{\perp0}&=-2\sqrt2N_B\sqrt{s_-}\,h_+\,\cT,\nonumber\\
A^T_{\parallel0}&=-2\sqrt2N_B\sqrt{s_+}\,
\widetilde h_+\,\cT,\nonumber\\
A^T_{\perp1}&=4N_B\sqrt{s_-}\frac{m_{\Lb}+m_{\Lc}}{\sqrt{q^2}}h_\perp\,\cT,\nonumber\\
A^T_{\parallel1}&=4N_B\sqrt{s_+}\frac{m_{\Lb}-m_{\Lc}}{\sqrt{q^2}}\widetilde h_\perp\,\cT.
 \label{eq:lbtensor}
\end{align}
The explicit factors of $1/m_\ell$ in the timelike amplitudes cancel against lepton-mass factors in the angular coefficients, yielding a finite massless limit.
The dictionary to Ref.~\cite{Boer:2019zmp} is
$(f_t^V,f_0^V,f_\perp^V)=(f_0,f_+,f_\perp)$,
$(f_t^A,f_0^A,f_\perp^A)=(g_0,g_+,g_\perp)$,
$(f_0^T,f_\perp^T)=(h_+,h_\perp)$, and
$(f_0^{T5},f_\perp^{T5})=(\widetilde h_+,\widetilde h_\perp)$.

The normalization of the baryonic angular coefficients follows by defining the vector--tensor interference term 
\begin{equation}
 {\cal T}=-2r_\ell\re\sum_{\lambda=\perp,\parallel}
 (A^T_{\lambda0}A_{\lambda0}^*
 +A^T_{\lambda1}A_{\lambda1}^*).\label{eq:vt-rate}
\end{equation}
The two coefficients needed to normalize ${\cal K}_{3s}$ are
\begin{align}
 {\cal K}_{1ss}={}&{\cal T}+\sum_{\lambda=\perp,\parallel}
 \bigg[\frac{1+r_\ell^2}{4}|A_{\lambda1}|^2
+\frac12|A_{\lambda0}|^2\nonumber\\
 &+\frac{r_\ell^2}{2}|A_{\lambda t}|^2
  +(1+r_\ell^2)|A^T_{\lambda1}|^2
+2r_\ell^2|A^T_{\lambda0}|^2\bigg],\label{eq:k1ss}\\
 {\cal K}_{1cc}={}&{\cal T}+\sum_{\lambda=\perp,\parallel}
 \bigg[\frac12|A_{\lambda1}|^2 +\frac{r_\ell^2}{2}|A_{\lambda0}|^2\nonumber\\
 &+\frac{r_\ell^2}{2}|A_{\lambda t}|^2
+2|A^T_{\lambda0}|^2+2r_\ell^2|A^T_{\lambda1}|^2\bigg].\label{eq:k1cc}
\end{align}
These are the unnormalized coefficients of Ref.~\cite{Boer:2019zmp}, and
\begin{equation}
 \frac{\dd\Gamma_B}{\dd q^2}=2{\cal K}_{1ss}+{\cal K}_{1cc}.
 \label{eq:baryonrate}
\end{equation}
Combining this rate with Eq.~\eqref{eq:k3s} gives $K_{3s}$ through Eq.~\eqref{eq:baryon-normalization} and its integrated moment over $m_\ell^2\le q^2\le(m_{\Lb}-m_{\Lc})^2$. The $q^2$ dependence of $N_B^2$ is retained in the integrated numerator and denominator.

\bibliographystyle{apsrev4-2}
\bibliography{baryonic}
\end{document}